%% file: main.tex
\documentclass[conference]{IEEEtran}
\IEEEoverridecommandlockouts
\usepackage{cite}
\usepackage{amsmath,amssymb,amsfonts}
\usepackage{algorithmic}
\usepackage{graphicx}
\usepackage{textcomp}
\usepackage{xcolor}
  \usepackage{cite}
\usepackage{cite}
\usepackage{amsmath,amssymb,amsfonts}
\usepackage{algorithmic}
\usepackage{graphicx}
\usepackage{textcomp}
\usepackage{comment}

\newtheorem{definition}{Definition}

\newtheorem{claim}{Claim}
\newtheorem{theorem}{Theorem}

\usepackage{soul}
\usepackage{xcolor}
\usepackage{color}

\usepackage{times}
\usepackage{amsfonts,amssymb}
\usepackage{multicol}
\newtheorem{observation}{\textbf{Observation}}

\renewcommand{\raggedright}{\leftskip=0pt \rightskip=0pt plus 0cm}
\usepackage{soul}
\usepackage{url}
\usepackage[hidelinks]{hyperref}
\def\done{\hspace*{\fill}\rule{1.8mm}{2.5mm}}
\usepackage[utf8]{inputenc}
\usepackage[small]{caption}
\usepackage{graphicx}
\usepackage{amsmath}
\allowdisplaybreaks[4]
\usepackage{booktabs}
 \usepackage{cancel} 
\usepackage{comment}
\usepackage{amsmath,lipsum}
\usepackage{cuted}
 \usepackage{algorithm}

\usepackage{amsmath}
\usepackage{amsfonts}
\usepackage{bbm}
\usepackage{dsfont}
\usepackage{diagbox} 
\usepackage{subfig}
\def\BibTeX{{\rm B\kern-.05em{\sc i\kern-.025em b}\kern-.08em
    T\kern-.1667em\lower.7ex\hbox{E}\kern-.125emX}}

\begin{document}

\title{Determining Blockchain Transaction Timing and Fee with observable \emph{mempools}
}

\author{Qianlan Bai, Yuedong Xu, Zhijian Zhou, Xin Wang
\thanks{Yuedong Xu is with College of Computer Science and Artificial Intelligence, and Artificial Intelligence Innovation and Incubation Institute, Fudan University, Shanghai, China (e-mail: ydxu@fudan.edu.cn)}
}

\maketitle

\input{section/abstract.tex}
\input{section/introduction.tex}
\input{section/measurement.tex}
\input{section/model_definition.tex}

\input{section/one_user_one_time.tex}

\input{section/multiple_markov}

\input{section/experiment.tex}

\input{section/related_work.tex}


\input{section/reference.tex}
\end{document}

%% file: section/abstract.tex
\begin{abstract}

Transaction fee plays an important role in determining the priority of transaction processing in public blockchain systems. 
Owing to the observability of unconfirmed transactions, a strategic user can postpone his transaction broadcasting time and set a fee as low as possible by prying into his \emph{mempool} that stores them. 
However, the stochastic mining interval may cause the delayed transaction to miss the next valid block. 
Meanwhile, a new feature (i.e. fee bumping) emerges that allows each user to increase his transaction fee before confirmation, making the fee setting more challenging. 
In this paper, we investigate a novel transaction policy from the perspective of a \emph{single} strategic user that determines the broadcasting time and the transaction fee simultaneously. Two representative scenarios are considered, in which a number of coexisting ordinary users are mempool-oblivious that set their fees according to certain distribution, and are semi-strategic that check their mempools at a Poisson rate and update their fees. In the former, we compute the optimal broadcasting time and transaction fee that adapts to the arbitrary distribution of mining interval. 
When the block interval is exponentially distributed in Bitcoin-like PoW systems, the strategic user needs to broadcast his transaction immediately after its creation. And when the block interval is fixed in Ethereum-like PoS systems, he finds it profitable to wait until the last moment before block generation. In the latter, we formulate a continuous-time Markov chain to characterize the dynamics of mempool states, and derive the optimal fee adjusting frequency of the strategic user when the block interval is exponentially distributed. In both theory and simulations, we show that this strategic user should immediately increase his fee whenever it falls behind the minimum fee of being included.

\end{abstract}

%% file: section/introduction.tex
\section{Introduction}


Blockchain technology has been paid great attention since the advent of Bitcoin in 2008  \cite{ref:bitcoin}.  A blockchain system operates as a decentralized ledger that records transaction history across a network of nodes and relies on its consensus mechanism to guarantee security.  The nodes who acquire the right to record the transactions are called \emph{miners}, and a miner generating a legitimate block is rewarded with a certain amount of X-coins from the system and transaction fees from the users.  Due to the limited capacity of each block to accommodate transactions, some influx of transactions surpassing this limit must wait to be included in future blocks. With the absolute or relative diminishing of X-coins in mainstream blockchain systems, the transaction fee becomes a crucial source of revenue for miners.  Miners prioritize high-yielding transactions, prompting user competition for limited block space which places the design and analysis of the transaction fee mechanism at the heart of blockchain technology \cite{ref:liuyunshuJSAC}.

The existing literature on blockchain transaction fees has two categories. One is the analysis of transaction confirmation delay of a \emph{tagged user} in the Bitcoin system, and the other is the prediction of the minimum transaction fee that this transaction can be included. 
 In the former, Kasahara \cite{ref:bernoulli exponential2} and Kawase et al. \cite{ref:priority queueing analysis} modeled the transaction processing as a priority queue with batch service, and computed the expected confirmation time 
of a transaction. Fiz et al. \cite{ref:confirmation delay prediction} adopted supervised learning to predict the probability of the successful inclusion for a given transaction fee. In the latter, 
AI-Shehabi \cite{ref:al-shehabi} and Shang et al. \cite{ref:need for speed} developed efficient machine-learning methods to help Bitcoin users estimate fees based on their preferred confirmation delay. Tedeschi et al. \cite{ref:tx fee distribution} designed a novel model of transaction inclusion and predicted it using a multi-layer neural network. A common feature of these two categories of studies is that every transaction is broadcasted to network immediately after its creation.

Blockchain allows everyone access to unconfirmed transaction data in the \emph{mempool}. Therefore, a \emph{strategic} user may surveil the mempool before deciding his transaction fee in the belief that observing adequate mempool information will grant him the chance of paying less to be included. He needs to broadcast his transaction at the right moment and set the transaction fee to the minimum amount of successful inclusion. However, two challenges hinder devising an optimal strategy. First, the block mining time interval is usually stochastic, which means that the ``lazy’’ broadcasting may risk losing the opportunity of sending out the transaction before the mining round terminates. Second, the introduction of fee bumping allows a user to increase his transaction fee one or more times before confirmation. These new features 
 complicate the fee setting and have not been studied to the best of our knowledge.

In this paper, we are particularly interested in the optimal transaction policy of a \emph{single} strategic user that is a two-tuple: the \emph{broadcasting time} and the \emph{amount of cryptocurrencies}, considering the randomness of block generating time and the fee bumping behavior. By focusing on a single strategic user, we are able to explore the achievable saving in today’s blockchain consensus and transaction mechanisms. As for other ordinary users, both the ``mempool-oblivious’’ scheme and the ``semi-strategic’’ scheme are investigated. In the former, an ordinary user does not pay attention to the mempool, and his fee is a random variable drawn from a certain distribution. In the latter, even (a part of) ordinary users may be smart to a certain extent so as to check the mempools and modify their fees from time to time. The fee adjustment rule of a semi-strategic user is relatively simple, that is, making it slightly higher than the minimum of transaction inclusion at each observation moment. We model such behaviors as a Poisson process with arrival rate as the frequency of possible fee bumpings. Our study is carried out from three aspects.

First, we measure the transaction patterns of Bitcoin and Ethereum to unveil the potential of strategic fee setting. The data analysis presents some interesting observations. i) By constructing a prophetic strategy, i.e. the minimum fee for successful inclusion, one can observe a huge transaction fee gap between this strategy and a referenced heuristic method. ii) By postponing broadcasting a transaction, more information regarding the mempool is acquired that might yield a higher revenue, and yet a higher probability of missing the broadcasting opportunity. These empirical insights derived from real data underscore the importance of devising a strategic policy.

Second, we formulate a stochastic model to capture the utility of the strategic user under different consensus mechanisms in which ordinary users are mempool-oblivious. The policy of the strategic user does not affect that of any ordinary user. We define three policies: NBR, IBR and FBR, where FBR is our focus. NBR is oblivious to the mempool, IBR considers the mempool but decides the transaction fee only and broadcasts immediately, and FBR decides both the broadcasting time and the transaction fee. The mining interval distribution influences the transaction broadcasting time where a special case (e.g. Bitcoin-like PoW systems) is to broadcast immediately due to the exponentially distributed, memoryless mining interval. At the other extreme, e.g. Ethereum-like PoS systems, the mining interval is fixed so that the optimal broadcasting time is just before the mining finishes.

Third, we develop a continuous-time Markov process to characterize the dynamic nature of mempool states, particularly when semi-strategic users engage in mempool checking and adopt fee bumping method. This simplified modeling approach captures their tactics through the Poisson arrival rate of fee bumping requests. The strategic user keeps observing the mempool and determines the optimal frequency of initiating his fee bumping actions. 
A notable observation in the Bitcoin-like system is that a strategic user's higher fee bumping frequency generates superior utility.
We present a concise formal proof that eliminates the influence of the number of user participants on this conclusion, primarily attributed to the distribution of block generating interval.

Our major contributions are briefly summarized as follows.
\begin{itemize}
\item We measure transaction patterns in Bitcoin and Ethereum, uncovering the potential of strategic transaction fee setting. The empirical insights reveal the importance of devising a transaction fee strategy considering block generating intervals and broadcasting time.
    \item We present a stochastic model to capture the utility of the strategic user when other users are mempool-oblivious. We prove theoretically that the strategic user should broadcast his transaction immediately when the block interval is exponentially distributed, and wait until the last moment when the block 
    interval is fixed. 

    \item When the block generating interval is exponentially distributed, we developed a continuous-time Markov process to
    model the utility of the strategic user when other users adopt the fee bumping method. Through numerical analysis, we demonstrate that the strategic user should increase his transaction fee as soon as his fee falls behind.

\end{itemize}


%% file: section/measurement.tex
\section{Background and Motivation}
\label{section:motivation}

\subsection{Basics of Consensus and Transaction Fee}

Our primary focus lies in the Proof-of-Work (PoW) consensus \cite{ref:bitcoin} employed by Bitcoin and the Proof-of-Stake (PoS) \cite{ref:posQA} 
consensus employed by Ethereum, where the block generation process and block generation time are different.

\emph{\textbf{Bitcoin:}}
PoW is the consensus mechanism in the Bitcoin system where valid blocks are generated by searching for a nonce that solves a cryptographic puzzle. Miners attempt nonces randomly until one satisfies the requirement. This process is modeled as a Bernoulli trial, with the number of trials needed following a geometric distribution \cite{ref:bernoulli exponential2}. 
After adding a new block to the blockchain, the miner receives cryptocurrencies as mining rewards, including the block reward and the transaction fee. The proportion of mining rewards from transaction fees is rising \cite{ref:blockchair} in BTC, which shows that transaction fees are becoming increasingly important.

\emph{\textbf{Ethereum:}} 
In September 2022, Ethereum switched its consensus mechanism to PoS (ETH2.0) \cite{ref:posQA}. With PoS, miners become validators by staking ETH or sending cryptocurrencies to staking wallets. A random algorithm selects a subset of miners to create and verify blocks. Unlike Bitcoin, ETH2.0's mining does not rely on computing power-based coin tossing. The interval between valid blocks is fixed at 12 seconds \cite{ref:ethereum12s}\cite{ref:etherscan}, not a random variable. After this transition, the miner generating the next valid block is solely compensated through transaction fees \cite{ref:posQA}, making transaction fees crucial for incentivizing Ethereum miners.

	\begin{figure}[!ht]
 \vspace{-0.3cm}
		\centering
		\includegraphics[width=0.49\textwidth]{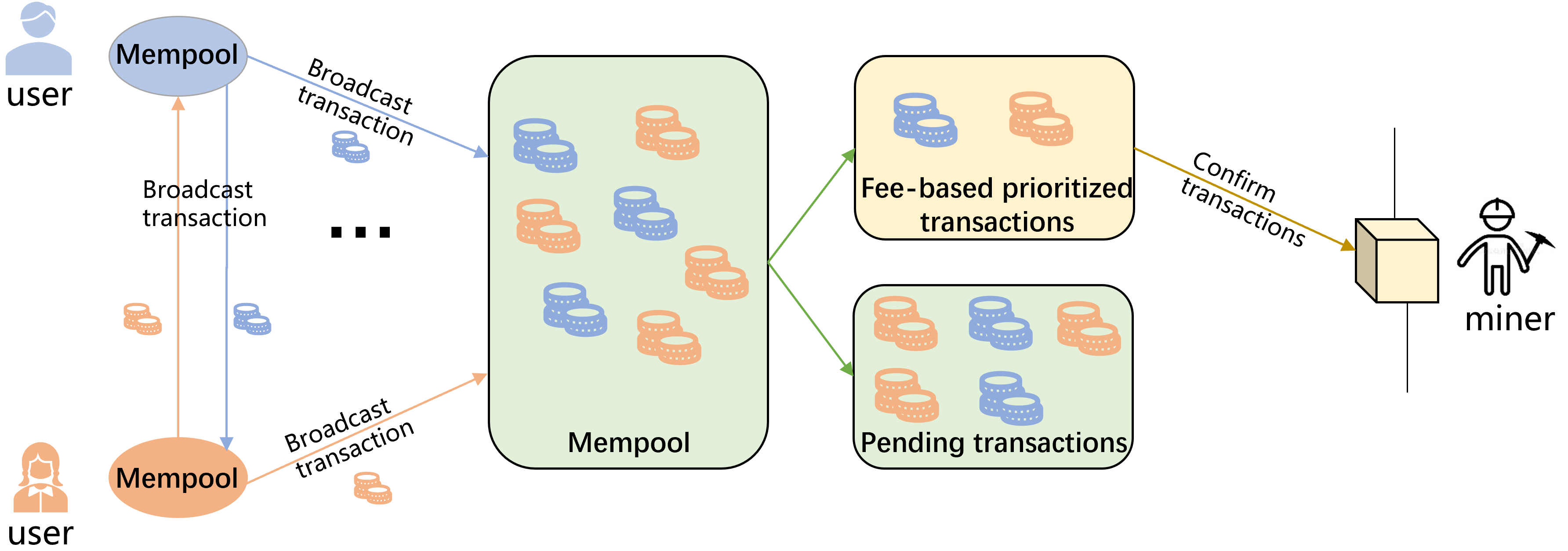}
    \vspace{-0.4cm}
		\caption{The transaction confirmation process.}
		\label{fig:tx confirmation process}
		\vspace{-0.3cm}
	\end{figure}

\textbf{Transaction confirmation process: }
The transaction confirmation process in Bitcoin and Ethereum (Fig. \ref{fig:tx confirmation process}) involves some key steps: a user initiates and broadcasts a transaction to the network of miners and users. Each transaction includes a transaction fee to incentive miners to confirm it. Each node independently verifies and adds transactions to the mempool, where all details are publicly accessible. Due to security concerns, the maximum number of transactions in a block is restricted \cite{ref:bitcoin} \cite{ref:eth gaslimit}. Miners follow a predetermined rule to rank and select transactions for the next block, prioritizing those with higher fees in practice. Unconfirmed transactions stay in the mempool and transfer to the subsequent mining round.

 \begin{figure}[htb]
 		\begin{minipage}[htb]{0.49\linewidth}
\setlength\abovecaptionskip{3pt}
\setlength\belowcaptionskip{-1pt}
\centering
		\includegraphics[width=0.9\textwidth]{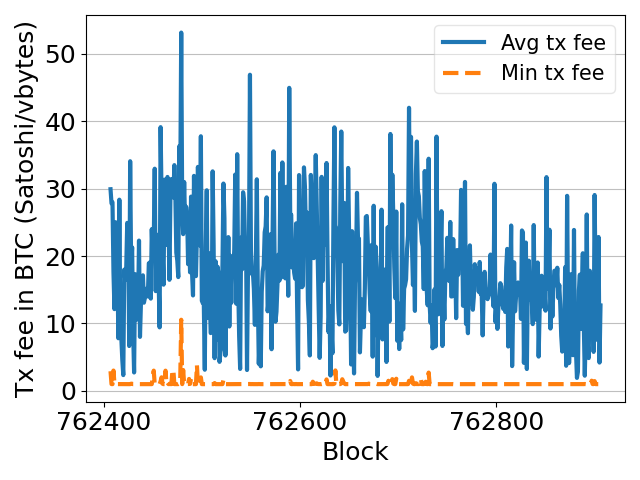}
		\caption{Average and minimum transaction fee in BTC system.}
		\label{fig:avg min tx fee btc}
\end{minipage}
 		\begin{minipage}[htb]{0.49\linewidth}
\setlength\abovecaptionskip{3pt}
\setlength\belowcaptionskip{-1pt}
\centering
		\includegraphics[width=0.9\textwidth]{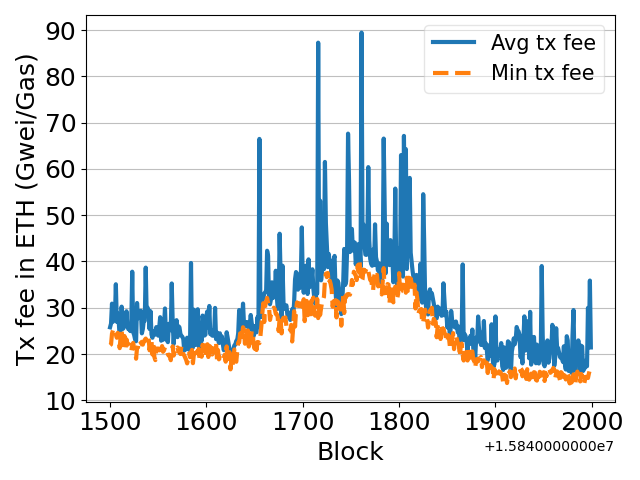}
    		\caption{Average and minimum transaction fee in ETH system.}
		\label{fig:avg min tx fee eth}
\end{minipage}
\vspace{-0.7cm}
\end{figure}

\subsection{Dilemma of Transaction Fee Setting}

Existing transaction fee recommendation algorithms in cryptocurrency wallets predominantly rely on historical data distribution \cite{ref:bitcoin core1} \cite{ref:metamask}\cite{ref:comparison}.
 To evaluate the current transaction fee strategy, we collect the transactions in BTC from Nov 28, 2020, to Oct 17, 2022. Fig. \ref{fig:avg min tx fee btc} and Fig. \ref{fig:avg min tx fee eth} show the average and minimum transaction fees per block in Bitcoin and Ethereum respectively. The noticeable difference between the minimum and average transaction fees suggests that users can strategically lower their transaction costs.

By judiciously utilizing mempool information, the strategic user can improve their transaction fee strategy. To investigate this hypothesis, we collect the transactions spanning blocks 762407 to 763020, including the confirmed transactions and unconfirmed ones, along with the corresponding transaction fees recommended by the Bitcoin core \cite{ref:statoshi} during this period. Each block contained an average of 1800 recorded transactions (in descending order). The transaction fee of the $1800^{th}$ ranked transaction is established as the threshold.

We record the mempool threshold values ($b$) at three time intervals after the last block's generation: 0.5$min$, 8$min$, and 15$min$. If the next valid block is generated before the scheduled time, we record it as the threshold at the last moment before block generation. With a budget of 30 Sats/vByte, we set transaction fees as $1.05b$ (Strategy 1), $1.15b$ (Strategy 2), and $1.25b$ (Strategy 3), ensuring they do not exceed the budget. We compare these transaction fees with the final threshold, and if the former is larger, the transaction is successfully included in the next block. The strategic user's utility is the difference between the budget and the transaction fee.

	\begin{table}[htb]
		\vspace{-0.2cm}
		\centering  
		\caption{Comparison of different transaction fees
 in BTC system.}
		{\begin{tabular}{|c|c|c|c|}
			\hline
			{}&{strategy1}&{strategy2}&{strategy3}\\
				\hline
    {0.5 min ($\star, \dagger$)}&{(52.69\%, 10.76)}&{(51.82\%, 11.79)}&{(66.39\%, 11.91)}\\
    \hline
    {8min ($\star, \dagger$)}&{(71.45\%,14.23)}&{(74.72\%,13.93)}&{(77.98\%,13.62)}\\
    \hline
    {15min ($\star, \dagger$)}&{(82.22\%,15.89)}&{(85.97\%,15.47)}&{(88.58\%,14.77)}\\
			\hline
   \multicolumn{4}{|c|}{ $\star$:success probability; $\dagger$:utility per vByte per block(Sats);}\\
   \hline
			\end{tabular}}
			\label{table:compare btc}
   \vspace{-0.3cm}
			\end{table}

Table \ref{table:compare btc} shows that as the time of obtaining $b$ becomes later, the success probability of exceeding the final threshold and utility of the strategic user increases. These utilities significantly surpass the utility of transaction fees based on the wallet's recommended price 
(10.08 Sats per vByte per block). The strategic user can indeed enhance their utility and improve the probability of successful processing by observing more transactions. However, the probability of missing the subsequent valid block increases as time elapses. The probabilities of the block generating interval being less than 0.5$min$, 8$min$, and 15$min$ are 5\%, 56\%, and 77\%, respectively. Therefore, there is a trade-off between observing more transactions and not missing the next valid block.

In light of the above observation, we are curious to know when a \emph{smart} user will broadcast his transaction and how much he will bid so as to reach the optimal tradeoff between the chance of being included in the forthcoming block and the paid transaction fee. 

%% file: section/model_definition.tex
\section{The Basic Model}
\label{sec:problem description}
\subsection{Problem Description}

 We present a novel transaction model of cryptocurrency systems with canonical settings in blockchain analysis \cite{ref:lijuanjuan1}\cite{ref:lijuanjuan2}.
We consider a transaction $i_t(b_i)\in \mathcal{N}$ submitted by user $i$ at time $t_i$ with the transaction fee $b_i$ and is broadcasted to other nodes at time $t$ ($t\geq t_i$). Each transaction can independently set the transaction fee and transaction broadcasting time. The valuation of each transaction is $V$ uniformly \cite{ref:liuyunshuJSAC}\cite{ref:from mining to market}.  
We use $G(T)$ to denote the distribution of block generating interval $T$, which is related to the consensus of the Blockchain system.
The distribution of the number of transactions $N_t$ at time $t$ is represented as $\Theta(N_t)$, and the state of a mempool at time $t$ is represented as $\mathbb{B}_t=\{b_1,\cdots, b_{N_t}\}$. We consider the mining of a brand-new block without unincluded residual transactions in the previous block.
Our emphasis is to explore how users can optimize their utilities by making use of the transaction information in the mempool. To this goal, we designate a specific user, referred to as the \emph{SP user} arriving at time $t_S$, and analyze its performance with transaction fee strategies and broadcasting time.
To facilitate the later analysis, we make the following assumptions on blockchain transactions.

\emph{Assumption 1: Each transaction has an equal size (resp. gas) in a size-limited (resp. gas-limited) system, and each block can process at most $m$ transactions \cite{ref:lijuanjuan1}\cite{ref:lijuanjuan2}\cite{ref:liuyunshuTON}.
}

Whilst the transaction size or gas varies in practice, large transactions can be split into small transactions with unit sizes. Therefore, this assumption has a negligible impact on the performance of the proposed model.

\emph{Assumption 2: The miner will select the top-m-fee-per-byte (resp. top-m-fee-per-gas) transactions in the mempool to include in the next valid block \cite{ref:liuyunshuJSAC}\cite{ref:lijuanjuan1}\cite{ref:lijuanjuan2}\cite{ref:liuyunshuicc}.
}

A rational miner will choose the transactions with high fees. Today's giant mining pools in Bitcoin such as ``AntPool'', ``F2Pool'' and ``BTC.com'' adopt the top-$m$-fee-per-byte scheme in a vast majority of cases (e.g. more than 90\%) \cite{ref:liuyunshuJSAC}.

Therefore, the utility of transaction \emph{SP} which is broadcasted at time $t$ can be given by:
\begin{align}
   \label{eq:utility}
U_S^t(b_S)=
\begin{cases}
& V-b_S \quad\text{if $b^{m}< b_S$ and $t\leq T$;}\\
&0\;\;\quad\quad\quad\text{otherwise.}
\end{cases}
\end{align}
Here, $b^{m}$ represents the lowest fee of a transaction which can be included in the next valid block. If less than $m$ transactions arrive at the mempool at time $T$, $b^{m}=0$.

\vspace{-0.3cm}
\subsection{Transaction Processing Timeline}

\subsubsection{Block generating interval}
 The block generating interval is usually determined by the blockchain consensus mechanism, in which PoW and PoS are investigated. 

\textbf{Bitcoin-like PoW.} 
The mining process involves a vast number of sequential Bernoulli trials conducted by miners. Each trial is independent and has a small success probability.
The block generating interval follows an exponential distribution, approximated by the extreme value theory \cite{ref:extreme}. And such an approximation has been well recognized in Bitcoin systems \cite{ref:block generation time1}\cite{ref:block generation time2}\cite{ref:gap game}. Hence, $G(T)$ can be written as
\begin{eqnarray}
\label{eq:exponential interval geo}
    G(T) =1-e^{-\lambda T},
\end{eqnarray}
where $\mathbb{E}[T]=\bar{T}=1/\lambda$. In different PoW systems, $\bar{T}$ varies considerable, i.e. 10 minutes in Bitcoin \cite{ref:bitcoin} and Bitcoin Cash\cite{ref:bitcoincash}, 2 minutes in Monero \cite{ref:monero}, and 15 seconds in ETC \cite{ref:ethereum classic}.

\textbf{Ethereum-like PoS.} The block generating time is usually fixed in Ethereum-like PoS consensus so that $T$ is deemed as a special discrete random variable with 
\begin{eqnarray}
G(T)=
\begin{cases}
& 1 \quad\text{if $T=\overline{T}$;}\\
&0\;\quad\text{otherwise.}
\end{cases} 
\end{eqnarray}
The value $\bar{T}$ is taken as 12 seconds in the Ethereum system.

\subsubsection{Transaction arrival process} 
Without loss of generality, we consider two arrival processes, the linear arrival process and the Poisson arrival process. 
The linear arrival is useful to approximate the arrival pattern for the blockchain systems with relatively short block generating intervals. The number of transactions in the mempool is $N_t=\beta t$ with $\beta$ denoted as the arrival rate per unit of time. The Poisson arrival is more often adopted in real-world systems that takes the following probability \cite{ref:bernoulli exponential2}\cite{ref:priority queueing analysis}\cite{ref:lijuanjuan queuing} 
\begin{eqnarray}
    \Theta(N_t=n)={(\beta t)^n}e^{-\beta t}/{n!}.
\end{eqnarray}

\subsubsection{Transaction fee bumping}
Bitcoin \cite{ref:RBF bitcoincore}\cite{ref:bitcoin core change limit}  and Ethereum \cite{ref:metamask adjust fee} enable users to pay additional transaction fees for existing transactions without re-signing or modifying the sequence. This new feature increases the likelihood of inclusion but lacks sufficient attention \cite{ref:BIP125 percentage}. In 2022, less than 30\% of Bitcoin transactions exhibit fungible signals, indicating that most users have not activated this option, and fee bumping has not been adopted in Bitcoin Cash \cite{ref:bitcoincash}. We envision that the fee bumping mechanism may intensify the severity of the competition. For completeness, both cases are considered: all users bid transactions once and all users can modify transaction fees.

\subsection{Other Users' strategies}

To examine the \emph{SP} user's optimal strategy, we specify the transaction fee bidding strategies of all the ordinary users.

\subsubsection{Mempool-oblivious Scheme}
The first strategy focuses on the condition that other users disregard the mempool state and fee bumping when setting transaction fees. 
In this case, the transaction fees of these users adhere to the distribution denoted as $F(b)$ with probability density function $f(b)$, which is frequently employed in relevant literatures \cite{ref:analysis of the confirmation time}\cite{ref:stablefees}\cite{ref:interplay tx security}. The time for the SP user to broadcast transactions only affects the number of observable transactions retrievable from the mempool. And the transaction fee set by the SP user does not affect the transaction fee setting of the ordinary users.

\subsubsection{Semi-strategic Scheme}
The second strategy under examination occurs within a system allowing transaction fee bumping. Each ordinary user randomly observes all transaction fees stored in the mempool. Upon discovering that his own transaction fee is lower than a threshold value $b^m$, he opts to increase his transaction fee to $b^m+\eta$. This evaluation process is a Poisson process with a rate $\gamma$ for the ordinary users (called the semi-strategic users). Meanwhile, the SP user continuously observes the mempool and increases his transaction fee to $b^m+\eta$ after a random interval when his fee is lower than the fee threshold. The random interval abides by an exponential distribution with a mean of $1/\gamma_S$. 
In contrast to the mempool-oblivious scheme, the SP user's fee and broadcasting time affect semi-strategic users' fee settings. As the SP user raises his fee at a higher rate, the threshold fee ($b^m$) also escalates. Semi-strategic users increase their fees accordingly. Lowering the  frequency of increasing fees may lead to potential omission of the subsequent valid block.

%% file: section/one_user_one_time.tex
\section{Best Strategy Against Mempool-oblivious Users}
\label{sec:bestresponse}

\subsection{Candidate Strategies}

We hereby define a suit of intuitive and reasonable strategies, while making the paid transaction fee or the broadcasting time as crucial variables to be optimized.

\begin{definition}
      \emph{Given the parameters of system $\Lambda=\{F(b),G(T),\Theta(n)\}$ and the current state of mempool $\mathbb{B}$, the SP user can take one of the following strategies.}
\begin{itemize}
    \item Naive Best-Response (NBR) Strategy: $b^N=\arg\max\limits_{b\leq V}\{\mathbb{E}[U_S^{t_S}(b)]|\Lambda\}$;
    \item Instant Best-Response (IBR) Strategy: $b^I=\arg\max\limits_{b\leq V}\{\mathbb{E}[U_S^{t_S}(b)]|\Lambda,\mathbb{B}\}$;
    \item Farsighted Best-Response (FBR) Strategy: $\{b^F,t^F\}=\arg\max\limits_{b\leq V,t\geq t_S}\{\mathbb{E}[U_S^{t}(b)]|\Lambda,\mathbb{B}\}$.   
\end{itemize}   
\end{definition}

The {NBR} strategy sets a fee based on the elapsed time since the creation of the previous block and promptly broadcasts the transaction to maximize utility, disregarding mempool information.
The SP user with IBR strategy considers mempool transactions to decide the fee but still broadcasts the transaction at time $t_S$ immediately.
The {FBR} strategy is a two-dimensional policy determining the transaction fee and broadcasting time. 
It assesses whether postponing broadcasting transactions yields a better utility for the SP user. Next, we explore the utilities of the SP user within different systems.

\subsection{Analysis of NBR strategy without fee bumping}

    The solution of NBR for SP user can be obtained by solving the following equations:
    \begin{align}
    &b^N=\arg\max_{b\leq V}\{R_N(b)\},\\
        & R_N(b)=\mathbb{E}[U_S^{t_S}(b)|\Lambda]=(V-b)\cdot W_N(b),
        \label{eq:NBR strategy}\\
    \label{eq:NBR success probability}
   & W_N(b)=    \sum\limits_{n=0}^{\infty}G(T|t_S)\Theta(n|T) \sum\limits_{j=0}^{m-1}C_n^j{F}(b)^{n-j}\overline{{F}}(b)^j, 
    \end{align}
    where $W_N(b)$ represents the probability that the SP user with transaction fee $b$ can be successfully included in the block, and $\overline{{F}}=1-F$.
    
The expected utility of SP user is determined by the system's consensus mechanism and the transaction's arrival mode, which collectively influence its specific form. 
When the transactions arrive linearly in the Bitcoin-like PoW system, Eq. \eqref{eq:NBR strategy} can be rewritten as:
\begin{align}
\label{eq:NBR simple utility pow}
    R_N(b)=&\frac{(V-b)}{q^{\lfloor t_S \beta \rfloor}}[1-(\frac{q-q{F}(b)}{1-q{F}(b)})^{m}-\\\nonumber
    &\sum\limits_{n=0}^{\lfloor t_S \beta \rfloor-1}(1-q)q^n\sum\limits_{j=0}^{m-1}C_n^j{F}(b)^{n-j}(\bar{F}(b))^j ],
\end{align}
where $q={\beta\bar{T}}/{(1+\beta\bar{T})}$.
If the transaction arrival process is a Poisson process with arrival rate $\beta$, there has 
\begin{align}
    R_N(b)=&(V-b)\sum\limits_{n=0}^{\infty}\frac{\lambda \beta^n}{(\lambda+\beta)^{n+1}}\\\nonumber
    &\sum\limits_{j=0}^{n}\frac{1}{e^{\beta t_S}}\frac{t_S^j(\lambda+\beta)^j}{j!} \sum\limits_{k=0}^{m-1}C_n^k{F}(b)^{n-k}\overline{{F}}(b)^k.
\end{align}

In the Ethereum-like PoS system, the block generation interval is usually fixed, then the expected utility of the SP user with fee $b$ expressed below. 
\begin{itemize}
    \item If the transaction arrive linearly with rate $\beta$, there is
    \begin{align}
    R_N(b)=
\begin{cases}
& V-b \quad\quad\quad\quad\quad\quad\quad\quad\quad\quad\quad\text{if $n< m$,}\\\nonumber
&(V-b)\sum\limits_{j=0}^{m-1}C_{n}^j{F}(b)^{n-j}\overline{{F}}(b)^j\;\text{otherwise,}
\end{cases}
    \end{align}
 where $n=\lfloor \beta\bar{T} \rfloor$.

\item If the transaction arrival process is Poisson with  rate $\beta$, there has
    \begin{align}
 &   R_N(b)=(v-b)\sum\limits_{n=0}^{\infty}e^{-\beta\bar{T}}\frac{(\beta\bar{T})^n}{n!}\sum\limits_{j=0}^{m-1}C_n^jF(b)^{n-j}\bar{F}(b)^j.\nonumber
    \end{align}
\end{itemize}

\subsection{Analysis of IBR strategy without fee bumping}
The solution of IBR strategy for SP user can be obtained by solving the following equations:
       \begin{align}
    &b^I=\arg\max_{b\leq V}\{R_I(b)\},\\
        & R_I(b)=\mathbb{E}[U_S^{t_S}(b)|\Lambda,\mathbb{B}]=(V-b)\cdot W_I(b),
    \label{eq:IBR strategy}\\
    &W_{I}(b)=\sum\limits_{n=0}^{\infty}G(T|t_S)\Theta(n|T) \sum\limits_{j=0}^{m'-1}C_n^j{F}(b)^{n-j}\overline{{F}}(b)^j,
    \label{eq:success probability IBR}
    \end{align}
    where $W_I(b)$ is the success probability and $O_{t_S}(b)$ indicates the number of users whose transaction fee are more than $b$ in the mempool at time $t_S$, and  $m'=m-O_{t_S}(b)$.

In Bitcoin-like PoW systems, Eq. \eqref{eq:IBR strategy} can be rewritten as:
\begin{align}
\label{eq:IBR simple utility pow}
    &R_I(b)=(V-b)[1-(\frac{q-qF(b)}{1-qF(b)})^{m-O_{t_S}(b)}].
\end{align}
If the transaction arrival process follows a Poisson process or linear arrival process with rate $\beta$, there has 
    $q={\beta}/{(\lambda+\beta)}$.

In the Ethereum-like PoS system, the expected utility of SP user when the transaction fee is $b$ can be rewritten as bellow.
\begin{itemize}
    \item If the transactions arrive linearly with rate $\beta$, then
    \begin{align}
    R_I(b)=
\begin{cases}
& V-b \quad\quad\quad\quad\quad\text{if $\beta \bar{T}< m$;}\\
&(V-b)W_I(b)\;\;\quad\text{otherwise.}\\
\end{cases}
\end{align}
\begin{align}
W_I(b)=\sum\limits_{j=0}^{m-O_{t_S}(b)-1}C_{\lfloor\beta (\bar{T}-t_S)\rfloor}^j{F}(b)^{\lfloor \beta (\bar{T}-t_S)\rfloor-j}\overline{{F}}(b)^j.\nonumber
    \end{align}
\item If the transaction arrival process is Poisson with rate $\beta$, there has
    \begin{align}
 &R_I(b)=(V-b)[\sum\limits_{n=0}^{\infty}e^{-\beta(\bar{T}-t_S)}\frac{[\beta(\bar{T}-t_S)]^n}{n!}\\\nonumber
 &\sum\limits_{j=0}^{m-O_{t_S}(b)-1}C_n^jF(b)^{n-j}\bar{F}(b)^j]. 
    \end{align}
\end{itemize}

We possess several important properties in the Bitcoin-like 
PoW system.
\begin{claim}[Properties of the IBR strategy in Bitcoin-like PoW system]
\emph{
\label{lemma:IBR properties}
\begin{itemize}
\item [i)] $W_I (b)$ is increasing in $b$ on $[0,V]$;
\item [iii)]$b^I$  will be non-decreasing with the increase of $V$;
\item [iv)] $R_I(b)$ will decrease with the increase of $\beta$. 
\end{itemize}}
\end{claim}

Here, (i) shows the strategic user can obtain higher success probability by paying more transaction fees.
(ii) shows the strategic user tends to set more transaction fee when $V$ is larger. According to a theoretical analysis,  (iii) shows that the increase of the transaction number will cause the decline in the utility of the strategic user.

\subsection{Analysis of FBR strategy without fee bumping}

    The solution of FBR for the SP user can be obtained by solving the following equations:
       \begin{align}
           \label{eq:FBR strategy}
    &(b^F,t^F)={\arg\max}_{b\leq V,t\geq t_S}\{R_F(b,t)\},\\
        & R_F(b,t)=\mathbb{E}[U_S^t(b)|\Lambda,\mathbb{B}]=(V-b)\cdot W_F(b,t),
    \end{align}
    where $W_{F}(b,t)$ indicates the success probability of the SP user to be included in the next valid block when he broadcasts the transaction at time $t$ and set the transaction fee as $b$.

Due to the exponential distribution is memoryless, the SP user has constant expectations of the number of upcoming transactions in Bitcoin-like PoW system. In Ethereum-like PoS system, the SP user can set the transaction fee after obtaining all other transactions' information. We formally state this conclusion in the following theorem.

\begin{theorem}
\label{theorem:FBR}
\emph{
    In the Bitcoin-like PoW system, the solution of FBR strategy for the SP user is $(b^F,t^F)=(b^I,t_S)$; in the Ethereum-like PoS system, the solution is $(b^F,t^F)=(b^m+\epsilon,\overline{T})$,
    where $\epsilon$ is the smallest unit of cryptocurrency.}
\end{theorem}

\textbf{Proof:}
First, we enumerate the case in the Bitcoin-like PoW system. 
Let the transaction fee of the SP user is $b$, and the utilities of broadcasting the transaction at time $t_S$ and the next time slot $t'$ are:
\begin{align}
    &W_F(b,t_S)=\sum\limits_{n=0}^{\infty}G(T|t_S)\Theta(n|T) \sum\limits_{j=0}^{m_1}C_n^j{F}(b)^{n-j}(\bar{F}(b))^j,\nonumber
    \\
      &W_F(b,t')=q\cdot \sum\limits_{n=0}^{\infty}G(T|t')\Theta(n|T)\sum\limits_{j=0}^{m_2}C_n^j F(b)^{n-j}(\bar{F}(b))^j.\nonumber
\end{align}
There has
\begin{align}
&G(T|t')=G(T|t_S),\\
   &m_1=(m-O_{t_S}(b)-1)\geq m_2=(m-O_{t'}(b)-1).
\end{align}
We can find that $R_F(b,t_S)$ is always no less than $R_F(b,t')$ for each $b$.
Thus a user can not gain more utility by postponing broadcasting the transaction in the Bitcoin-like PoW system for any transaction fee.

We next prove the optimal solution of FBR strategy in the Ethereum-like PoS system. 
If the SP user broadcasts the transaction at time $\bar{T}$, he selects a transaction fee slightly higher than $b^m$ to ensure its inclusion in the block. The expected utility at time $t$ is
\begin{align}
    R_F(b,\overline{T})=\int_{0}^{V}(V-x-\epsilon)dW_F(x,t).
\end{align}
The winning probability function is not continuous in the entire feasible region, but continuous in the interval $[\underline{b},\overline{b}]$ when $O_{t}(\underline{b})=O_{t}(\overline{b})$.
In order to prove that $R_F(b,t)$ is always no more than $R_F(b^F,\overline{T})$, we will set
\begin{align}
    &D=(V-b)W_F(b,t)-\sum\limits_{j=1}^{k}\int_{\underline{b_j}}^{\overline{b_{j}}}(V-x-\epsilon)dW_{F}(x,t)dx,
    \label{eq:D}\nonumber\\
    &k=\max\{m-O_{t}(b), b\leq V\}\leq m,\\
       &\underline{b_j}=\arg\min_{b\in \mathbb{B}_{t}}\{O_{t}(b)=j\},\\\nonumber
       &\overline{b_j}=\arg\min_{b\in \mathbb{B}_{t}}\{O_{t}(b)=j+1\},\\\nonumber
       &b\in \{[\underline{b_1},\overline{b_1})\cup[\underline{b_2},\overline{b_2})\cup\cdots\cup[\underline{b_k},\overline{b_k}]\}.
\end{align}
Considering $V\leq \overline{b_k}$ and $\epsilon$ is so small that can be ignored in current cryptocurrency system, we can rewrite $D$ as:
\begin{align}
    D=&(V-b)W_F(b,t)-\sum\limits_{j=1}^{k-1}\int_{\underline{b_j}}^{\overline{b_{j}}}(V-x)dW_F(x,t)-\\\nonumber
    &\int_{\underline{b_k}}^{V}(V-x)dW_F(x,t).
\end{align}
Taking the derivative of $D$ with respect to $V$, we can obtain
\begin{align}
   \frac{d D}{d V}=W_F(b,t)-W_F(V,t)\leq 0. 
\end{align}
The maximum value of $D$ is $0$ when $V=b$. We can obtain that for each transaction fee, there is $D\leq 0$. Thus the expected utility of broadcasting transaction at time $\bar{T}$ will be no less than that of any other time. 
 \done

%% file: section/multiple_markov.tex
\section{Best strategy against semi-strategic
users}
\label{sec:continuous}

\subsection{Problem description}

The semi-strategic scheme mimics the situation that all the transactions users pay their transaction fees based on the mempool state. 
To the best of our knowledge, no known optimal strategy exists in the auction that a large number of users compete for many vacancies with multiple fee setting times in theory. Therefore, the semi-strategic scheme captures the wisdom of the other users appropriately and considers the mathematical tractability meanwhile. 
When other users are semi-strategic, the transaction strategy of SP user and other users' fee setting influence each other. This is because other users randomly observe the states of the mempool at a rate $\gamma$ and raise the transaction fee if it falls below the transaction fee threshold $b^m$. If the frequency $\gamma_S$ at which the SP user chooses to increase the transaction fee is excessively high, it results in inflated transaction prices and diminished average utilities. Consequently, our analysis will concentrate on the impact of the frequency of fee modification in a quasi-static state where the number of users does not change during one mining round \cite{ref:liuyunshuJSAC} \cite{ref:lijuanjuan2}\cite{ref:from mining to market}\cite{ref:liuyunshuicc}.

To facilitate the later analysis, we consider there are $n$ pending transactions in each block generation round ($n>m$). The SP transaction is noted as transaction $n$. We normalize the amount of each fee bumping $\eta$ to 1 and denote $V/\eta$ as $\hat{V}$.
To break the tie resulting from identical transaction fees, we posit that later transaction fees should offer a higher value of $\epsilon$ than earlier transaction fees, even though the transaction fee threshold remains unchanged. This minimal increment in value will have a negligible impact on user expenditures. However, it will effectively prioritize later fees for miners when multiple users place the same transaction fees.
In the Ethereum-like PoS system, a rational SP user will always broadcast his transaction at time $\overline{T}$ and set the transaction fee as $b^m+\epsilon$. Therefore, we mainly focus on the performance of the SP user when the 
block generating interval is exponentially 
distributed.

 \subsection{Continuous-time Markov models}
The block generation process and each user taking an action are Poisson processes so that the transaction fee competition can be modeled as a continuous time Markov process. We designate the state of the system as $X_t$, which belongs to the set $\mathcal{X}=\{0,Q(k,b,I),S(k,b,I):1\leq k \leq m, 1\leq b \leq \hat{V}, 0\leq I \leq m\}$. In this notation, $b$ denotes the current highest transaction fee in the mempool, $k$ represents the number of transactions with the transaction fee $b$, $I$ indicates the number of newly arrived transaction fees after the SP user's latest fee setting, $Q$ signifies the next block that has not been generated, and $S$ represents the next block that has been generated and this mining round is over. The stationary probabilities of each state $X$ are denoted as $\pi(x)=\lim_{t\to \infty}P(X_t=x)$.

The state transitions in the stationary Markov process occur at some time from the beginning of the mining as follows:
\begin{itemize}
    \item The mempool receives a new fee from the SP user with independent and exponentially distributed increase interval in the state $Q(k,b,I)$. In this case, if $(k < m)$, then the state transfers 
    to state $Q(k+1,b,0)$; if $(k=m, b<\hat{V})$, then the state transfers to $Q(1,b+1,0)$.
    \item Similarly, the mempool receives a new fee from other users with independent and exponentially distributed increase interval in the state $Q(k,b,I)$. In this case, if $(k<m,I<m)$, then the state transfers to $Q(k+1,b,I+1)$; if $(k=m, b<\hat{V}, I<m)$, then the state transfers to $Q(1,b+1,I+1)$;
    if $(k<m,I=m)$, then the state transfers to $Q(k+1,b,I)$;
    if $(k=m, b<\hat{V}, I=m)$, then the state transfers to $Q(1,b+1,I)$.
     \item After the next valid block is generated, the state will transfer to state $S(k,b,I)$ from state  $Q(k,b,I)$.
    \item If there is no users that have broadcast transactions to the network when the valid block is generated, the state transitions to state 0. 
\end{itemize}

The state transitions between all stable states can be represented as Eq. \eqref{eq:markov1} $\sim$ Eq. \eqref{eq:markov2}. 

\vspace{-0.8cm}
\begin{align}
\label{eq:markov1}
&\pi(0)[(n-1)\gamma+\gamma_S]=\lambda \sum_{k=1}^{n}\sum_{b=1}^{\hat{V}}\sum_{I=0}^{b}S(k,b,I);\\
&\pi(Q(1,1,0))[(n-1)\gamma+\lambda]=\\\nonumber &[\pi(0)+\sum_{k=1}^{n}\sum_{b=1}^{\hat{V}}\sum_{I=0}^{b}S(k,b,I)]\gamma_S;\\
&\pi(Q(1,1,1))[(n-2)\gamma+\gamma_S+\lambda]=\\\nonumber &[\pi(0)+\sum_{k=1}^{n}\sum_{b=1}^{\hat{V}}\sum_{I=0}^{b}S(k,b,I)](n-1)\gamma;\\
&\pi(Q(k,1,I))[(n-k)\gamma+\lambda]=\pi(Q(k-1,1,I-1))\\\nonumber &(n-k+1)\gamma, 1\leq I<k,1<k\leq m;\\
&\pi(Q(k,1,k))[(n-k-1)\gamma+\gamma_S+\lambda]=\\\nonumber &\pi(Q(k-1,1,k-1))(n-k)\gamma,1<k\leq m;\\
&\pi(Q(k,1,0))[(n-k)\gamma+\lambda]=\pi(Q(k-1,1,k-1))\gamma_S;\\
&\pi(Q(k,b,I))[(n-m)\gamma+\lambda]=\pi(Q(k-1,b,I-1))\\\nonumber &(n-m)\gamma,1< k \leq m, 1< b \leq \hat{V},0<I<m;\\
&\pi(Q(k,b,m))[(n-1-m)\gamma+\gamma_S+\lambda]=\pi(Q(k-1,b,m))\nonumber \\
&(n-1-m)\gamma+\pi(Q(k-1,b,m-1))(n-m)\gamma;\\
&\pi(Q(k,b,0))[(n-m)\gamma+\lambda]=\pi(Q(k-1,b,m))\gamma_S;\\
&\pi(Q(1,b,I))[(n-m)\gamma+\lambda]=\pi(Q(m,b-1,I-1))\\\nonumber &[(n-m)\gamma], 1<b \leq \hat{V}, 0<I<m,\\
&\pi(Q(1,b,m))[(n-1-m)\gamma+\gamma_S+\lambda]=\pi(Q(m,b-1,m))\\\nonumber
&(n-1-m)\gamma+\pi(Q(m,b-1,m-1))(n-m)\gamma,\\
&\pi(Q(1,b,0))[(n-m)\gamma+\lambda]=\pi(Q(m,b-1,m))\gamma_S, \\
&\pi(Q(m,\hat{V},I))\lambda=\pi(Q(m-1,\hat{V},I-1))[(n-m)\gamma],\nonumber\\
&0<I<m\\
&\pi(Q(m,\hat{V},m))\lambda=\pi(Q(m-1,\hat{V},m-1))(n-m)\gamma+\\\nonumber &\pi(Q(m-1,\hat{V},m))(n-1-m)\gamma,\\
&\pi(Q(m,\hat{V},0))\lambda=\pi(Q(m-1,\hat{V},m))\gamma_S,\\
&\pi(S(k,b,I))[(n-1)\gamma+\gamma_S+\lambda]=\pi(Q(k,b,I))\lambda.\\
&\pi(0)+\sum_{k=1}^{n}\sum_{b=1}^{\hat{V}}\sum_{I=0}^{m}[\pi[S(k,b,I)]+\pi[Q(k,b,I)]]=1. 
\label{eq:markov2}
\vspace{-0.3cm}
\end{align}

The expected utility can be expressed as 
\begin{align}
    &\mathbb{E}(U_i)={(\eta\cdot Y)}/({\sum\limits_{b=1}^{\hat{V}}\sum\limits_{k=1}^{m}\sum\limits_{I=0}^{m}\pi[S(k,b,I)]+\pi(0)}),\\ 
    &Y=\sum_{b=2}^{\hat{V}}\sum_{k=1}^{m}[\sum_{I=0}^{k-1}(\hat{V}-b)\pi(S(k,b,I))+\sum_{I=k}^{m-1}\pi(S(k,b,I))\nonumber\\
    &(\hat{V}-b+1)]+\sum_{k=1}^{m}\sum_{I=0}^{k-1}(\hat{V}-1)\pi(S(k,1,I)).
\end{align}

\begin{theorem}
    \emph{If $n=2, m=1$ and the block generating interval is exponentially distributed, the optimal strategy for the SP user is to set his next transaction fee as quickly as possible when other users are semi-strategic.} 
\end{theorem}

\textbf{Proof:} When $n=2$ and $m=1$, we assume the SP user will set the $i^{th}$ transaction fee after a time interval $t_1^i$ when his transaction fee is not the highest. And the ordinary user will set his transaction fee after interval $t_2^i$ ($H_i=t_1^i+t_2^i$) when his fee is not the highest. The SP user will win this competition only when the block is generated after he sets the transaction fee and before the other user sets the new transaction fee. If the SP user sets first, then the expected utility is
\begin{align}
    &\mathbb{E}[U]=0 \varpi_1+(\hat{V}-1)\varpi_2+
     0\cdot \varpi_3+(\hat{V}-3)\varpi_4+\cdots,\nonumber\\
    &\varpi_1=G(t_1^1),\nonumber\\
    &\varpi_2=G(t_1^1+t_2^1)-G(t_1^1),\nonumber\\
    &\varpi_{2r+1}=G(\sum_{j=1}^rH_j+t_1^{r+1})-
G(\sum_{j=1}^rH_j),\forall 1\leq r \leq \hat{V}/2,\nonumber\\
    &\varpi_{2r}=G(\sum_{j=1}^{r}H_j)-G(\sum_{j=1}^{r-1}H_j+t_1^{r}),\forall 1\leq r \leq \frac{\hat{V}+1}{2}.\nonumber
    \vspace{-0.5cm}
\end{align}
Take the derivative of the success probability of the SP user with respect to each $t_1^i$, we will find the derivatives are all less than 0. That means the SP user should minimize each $t_1^i$.
\done

In the experiment section, we numerically solve the above continuous-time Markov model to discuss the optimal strategy of SP user when $m>1$ and $n>2$.

%% file: section/experiment.tex
\section{Experimental study}
\label{sec:evaluation}

In this section, we evaluate the performance of the proposed strategies and verify our theoretical results. 
To align with the current gas limit in Ethereum, a single block can accommodate around 200 transactions \cite{ref:liuyunshuJSAC}\cite{ref:dynamic posted price}.
 The block generation rate $\lambda$ is set as $0.1$ and the transaction arrival rate $\beta$ is set as $40$ \cite{ref:betaref}.

\subsection{Experimental results of mempool-oblivious seniors}
 We analyze the Ethereum transactions on April 18, 2023 and observe an average fee of $5.9512milliether$ \cite{ref:etherscan}  
per transaction.
Therefore, the transaction fee distribution of other mempool-oblivious users is set as Pareto distribution \cite{ref:stablefees}\cite{ref:dynamic posted price}\cite{ref:stackelberg attack}\cite{ref:honeymoon} with a mean of $5.9512$
    and a minimum of $1$. 
In terms of the baseline, we choose the average transaction fee algorithm (average-strategy) that is widely used in blockchain wallets, and simplify it as the selection of $m^{th}$ value\cite{ref:bitcoin core1}\cite{ref:metamask}.

\begin{observation}
    \emph{The utility of IBR strategy decreases (resp. increases) and the utility of NBR strategy decreases (resp. remains constant) with the increase of elapsed time since the last valid block when the block generating interval is exponentially distributed (resp. fixed).} 
\end{observation}

\begin{figure}[htb]
		\begin{minipage}[htb]{0.49\linewidth}
			\setlength\abovecaptionskip{0.5pt}
			\setlength\belowcaptionskip{-1pt}
			\centering
			\includegraphics[width=0.9\textwidth]{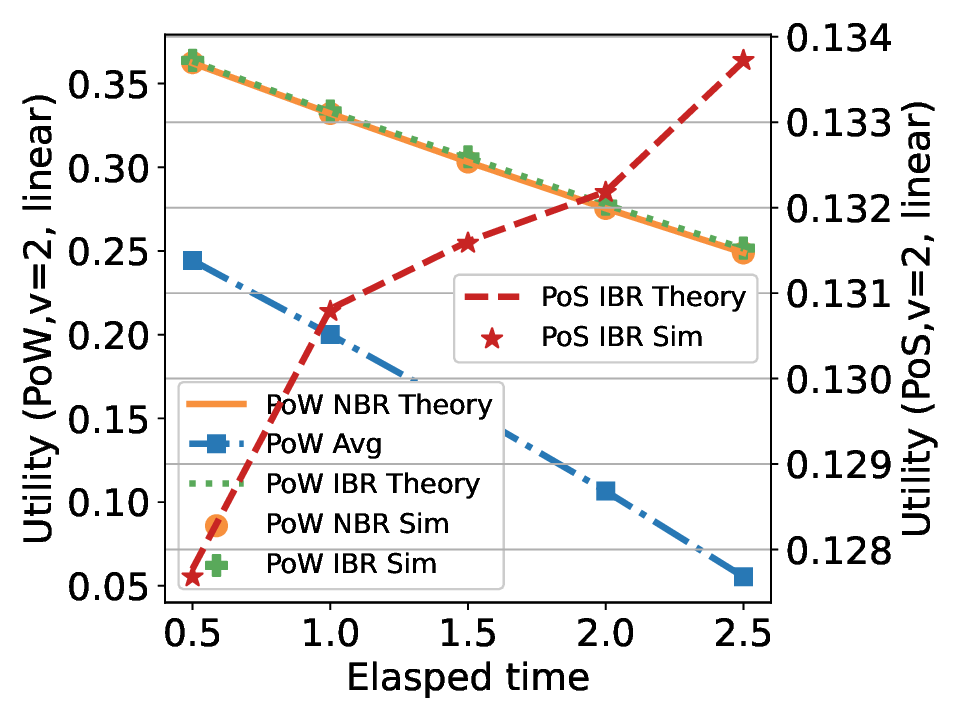}\\
			\caption{Utility of NBR and IBR with linear arrival when $V=2$.}
			\label{fig:NBRIBRlinearv2}
		\end{minipage}
  		\begin{minipage}[htb]{0.49\linewidth}
			\setlength\abovecaptionskip{0.5pt}
			\setlength\belowcaptionskip{-1pt}
			\centering
			\includegraphics[width=0.9\textwidth]{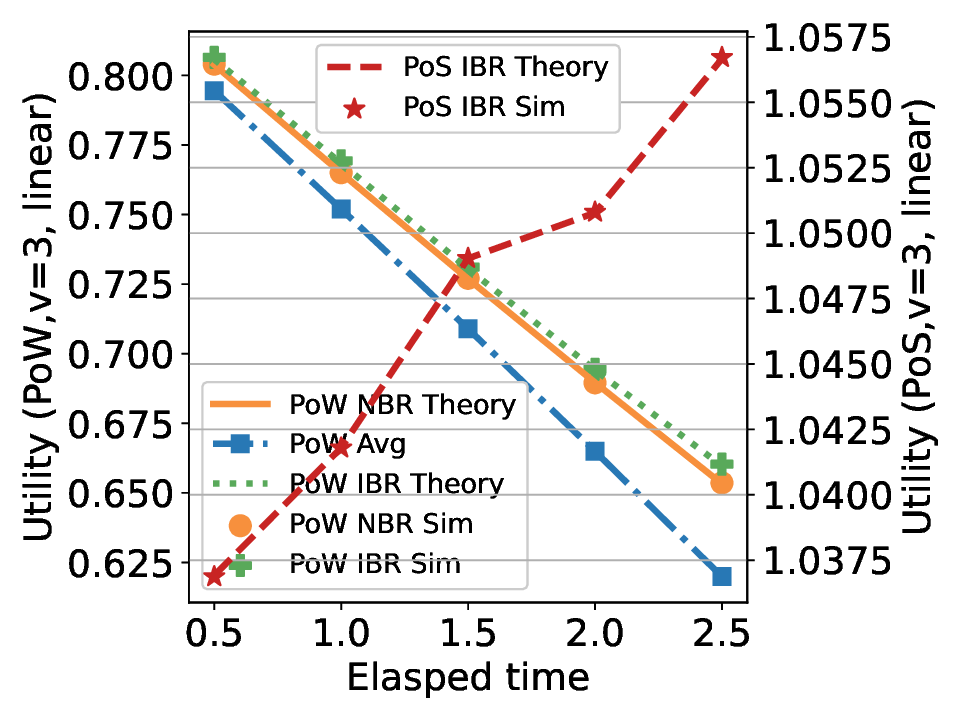}\\
			\caption{Utility of NBR and IBR with linear arrival when $V=3$.}
			\label{fig:NBRIBRlinearv3}

		\end{minipage}
    		\begin{minipage}[htb]{0.49\linewidth}
			\setlength\abovecaptionskip{0.5pt}
			\setlength\belowcaptionskip{-1pt}
			\centering
			\includegraphics[width=0.9\textwidth]{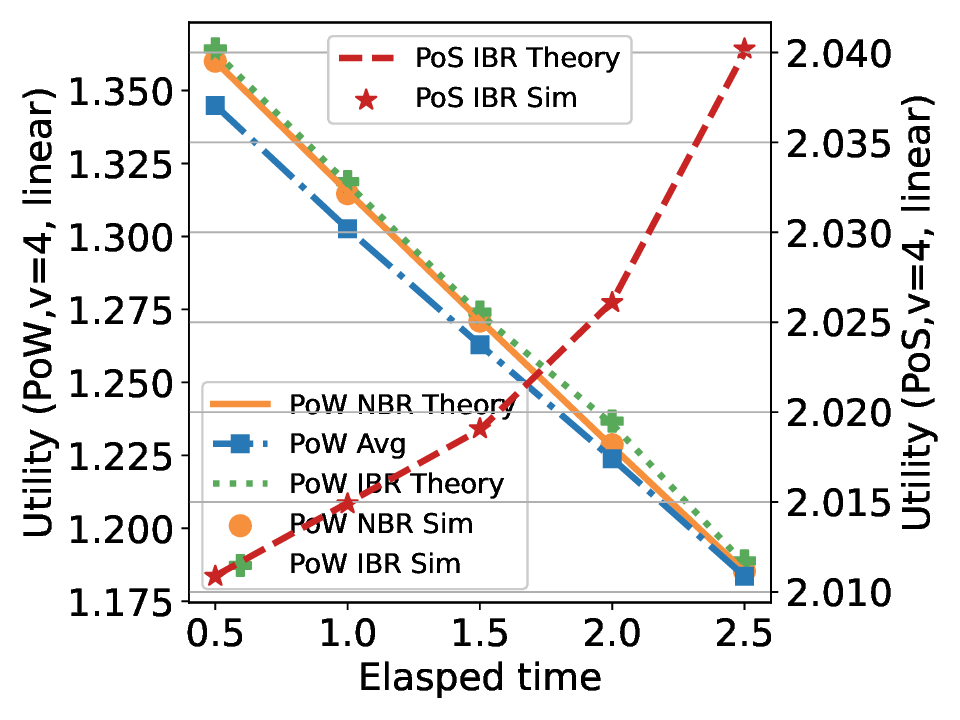}\\
			\caption{Utility of NBR and IBR with linear arrival when $V=4$.}
			\label{fig:NBRIBRlinearv4}
		\end{minipage}
		\begin{minipage}[htb]{0.49\linewidth}
			\setlength\abovecaptionskip{0.5pt}
			\setlength\belowcaptionskip{-1pt}
			\centering
			\includegraphics[width=0.9\textwidth]{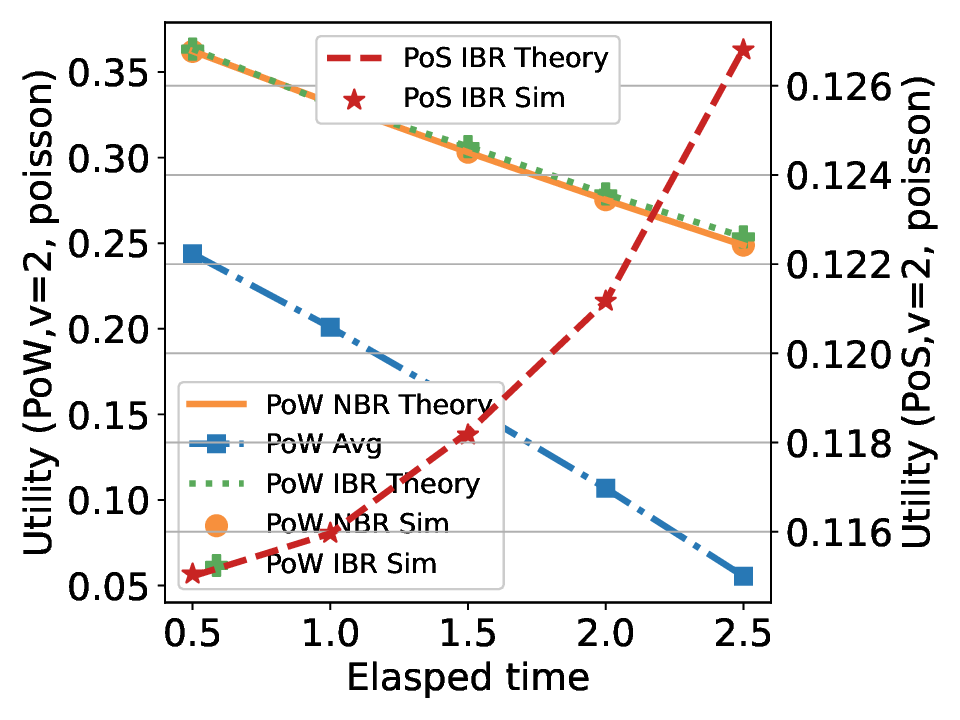}\\
			\caption{Utility of NBR and IBR with poisson arrival when $V=2$.}
			\label{fig:NBRIBRpoissonv2}
		\end{minipage}
  		\begin{minipage}[htb]{0.49\linewidth}
			\setlength\abovecaptionskip{0.5pt}
			\setlength\belowcaptionskip{-1pt}
			\centering
			\includegraphics[width=0.9\textwidth]{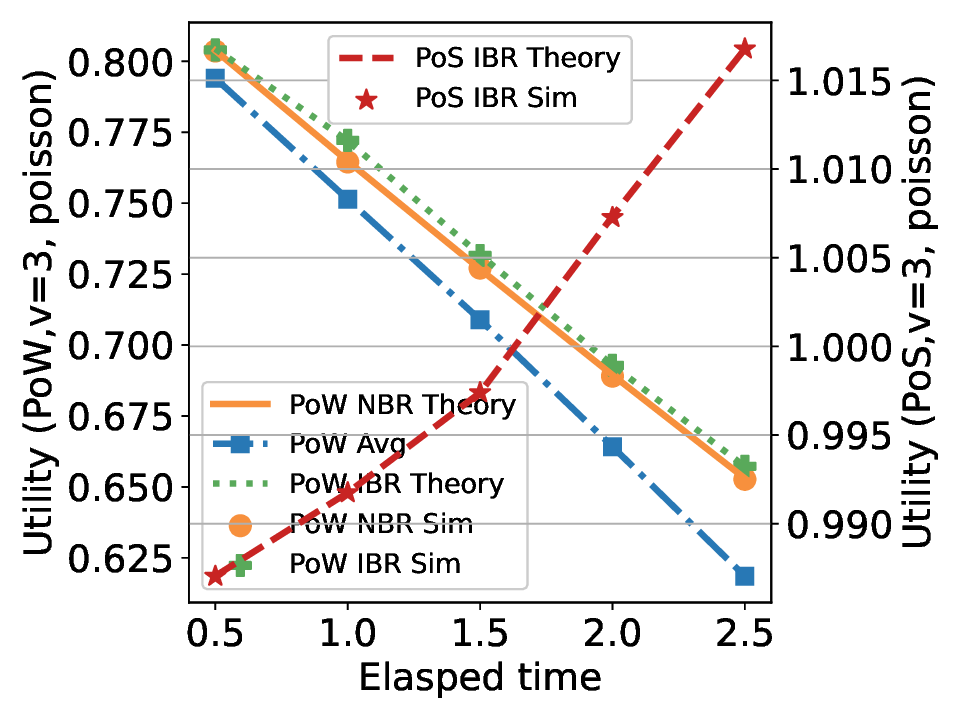}\\
			\caption{Utility of NBR and IBR with poisson arrival when $V=3$.}
			\label{fig:NBRIBRpoissonv3}
		\end{minipage}
    		\begin{minipage}[htb]{0.49\linewidth}
			\setlength\abovecaptionskip{0.5pt}
			\setlength\belowcaptionskip{-1pt}
			\centering
			\includegraphics[width=0.9\textwidth]{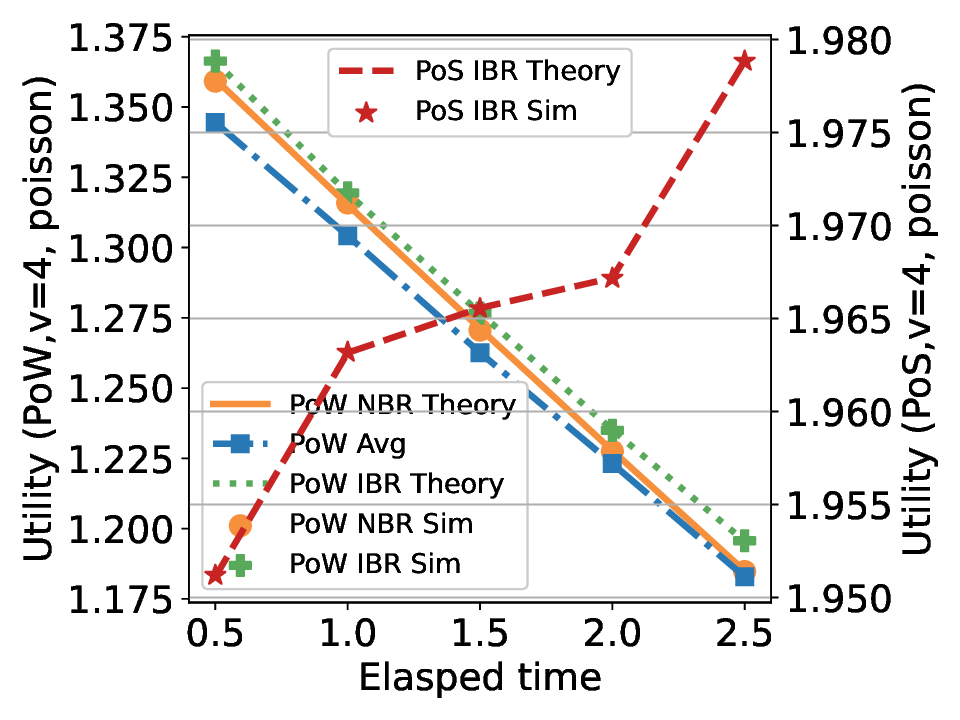}\\
			\caption{Utility of NBR and IBR with poisson arrival when $V=4$.}
			\label{fig:NBRIBRpoissonv4}
		\end{minipage}
  \vspace{-0.7cm}
	\end{figure}

We thereby show the utilities of the SP user with NBR and IBR strategy in Fig. \ref{fig:NBRIBRlinearv2}$\sim$Fig. \ref{fig:NBRIBRpoissonv4} under different values and different transaction arrival models. The figures consist of markers that depict simulation results, and lines that represent model results. The horizontal axis denotes the time elapsed since the most recent valid block (referred to as the elapsed time), while the vertical axis indicates the utility of the SP user. Different transaction values are selected for comparative analysis, revealing a strong consistency between the simulation results and theoretical outcomes across diverse scenarios.

Regardless of the value of the SP user, the results consistently demonstrate that utilities from NBR and IBR strategies exceed the utility obtained from average-strategy. In the Bitcoin-like PoW system, the utility of IBR strategy is slightly higher to the SP user than the NBR strategy, which means that the judicious use of the transaction information in the mempool is beneficial to the SP user. In the Ethereum-like PoS system, for linear transaction arrival, the utility of NBR strategy is (0.124, 1.032, 2.007) when the value is (2,3,4), while the utility of average-strategy is (0.1115, 0.6293, 1.146). In the case of a Poisson arrival, the utility of NBR strategy is (0.114, 0.980, 1.944), and the utility of average-strategy is (0.111, 0.624, 1.143). In all cases, the utilities of both the average-strategy and NBR strategy in the Ethereum-like PoS system are consistently lower than that of IBR strategy.

The utility of the NBR and IBR strategy in the Bitcoin-like PoW system decreases as the elapsed time since the last valid block generates increases. While in the Ethereum-like PoS system, the utility of IBR strategy increases with the elapsed time increases.
We speculate that this is due to the fact that the expected block interval increases as elapsed time increases in the Bitcoin-like PoW system, while it remains unchanged in the Ethereum-like PoS system. The expected increase in the block generation interval leads to a higher number of expected pending transactions. Consequently, the utility obtained by the SP user in the Bitcoin-like PoW system decreases, although more transactions can be observed from the mempool.

\begin{figure}[h]
 \vspace{-0.3cm}
   		\begin{minipage}[htb]{0.49\linewidth}
			\setlength\abovecaptionskip{0.5pt}
			\setlength\belowcaptionskip{-1pt}
			\centering
			\includegraphics[width=0.9\textwidth]{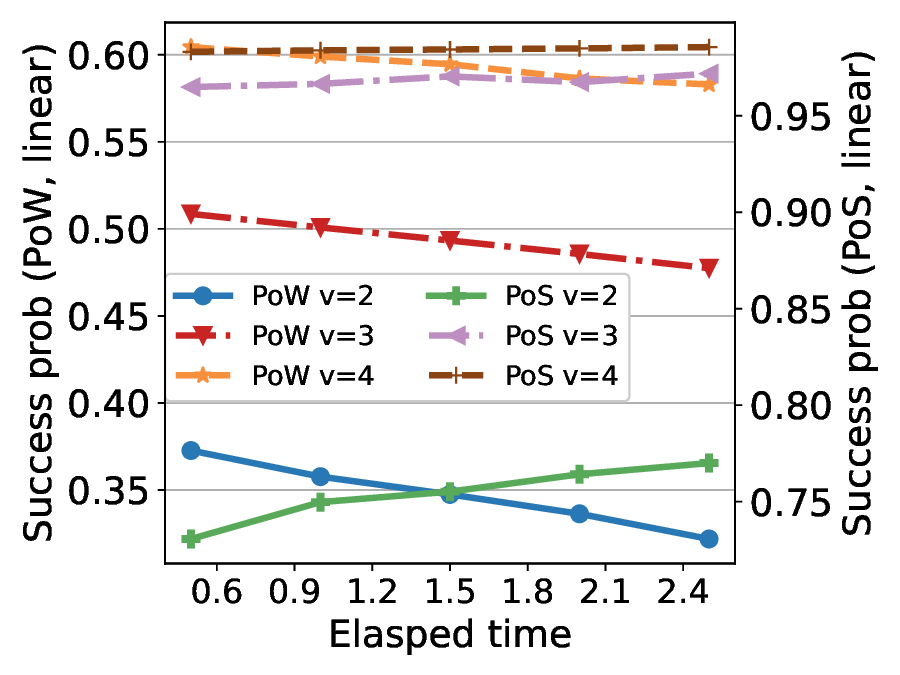}\\
			\caption{Probability of the SP user can be included with linear transaction arrival.}
			\label{fig:successlinear}
		\end{minipage}
  		\begin{minipage}[htb]{0.49\linewidth}
			\setlength\abovecaptionskip{0.5pt}
			\setlength\belowcaptionskip{-1pt}
			\centering
			\includegraphics[width=0.9\textwidth]{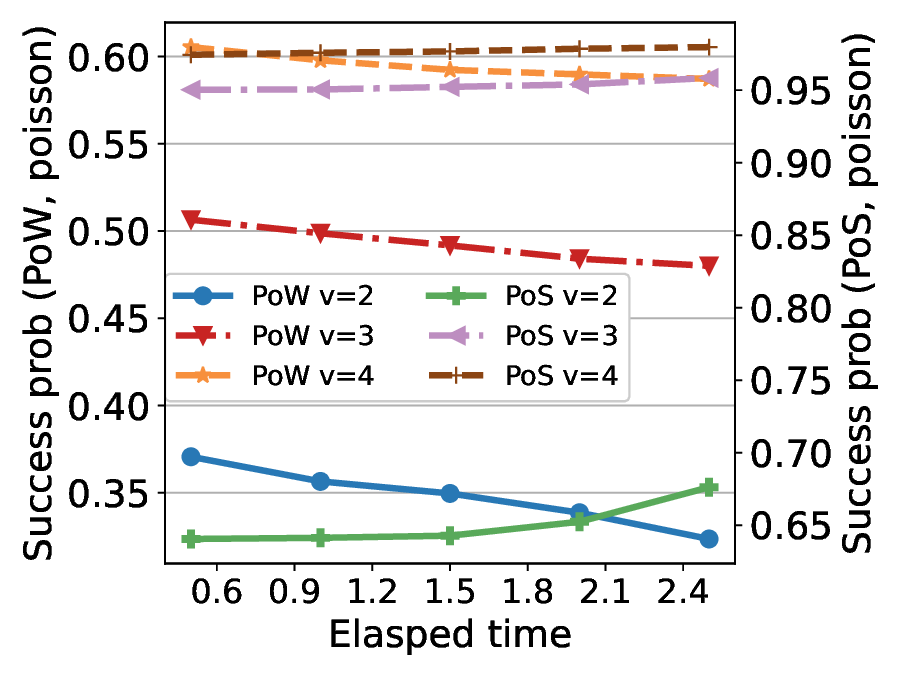}\\
			\caption{Probability of the SP user can be included with Poisson transaction arrival.}
			\label{fig:successpoisson}
		\end{minipage}
		\begin{minipage}[t]{0.49\linewidth}
			\setlength\abovecaptionskip{0.5pt}
			\setlength\belowcaptionskip{-1pt}
			\centering
			\includegraphics[width=0.9\textwidth]{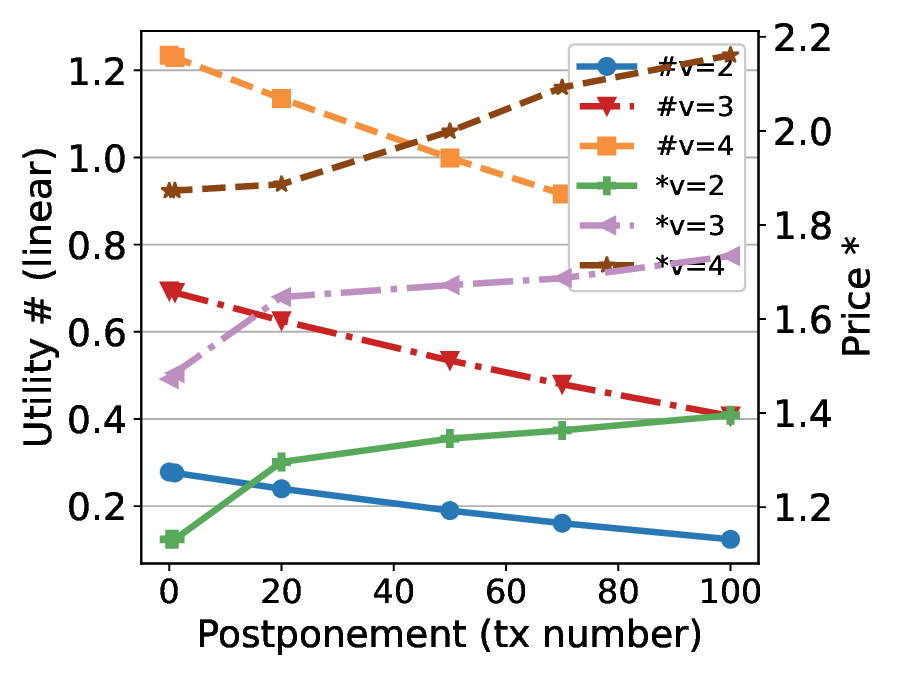}\\
			\caption{Utility and transaction fee of FBR in different postponements with linear arrival.}
			\label{fig:FBRlinear}
		\end{minipage}
		\begin{minipage}[t]{0.49\linewidth}
			\setlength\abovecaptionskip{0.5pt}
			\setlength\belowcaptionskip{-1pt}
			\centering
			\includegraphics[width=0.9\textwidth]{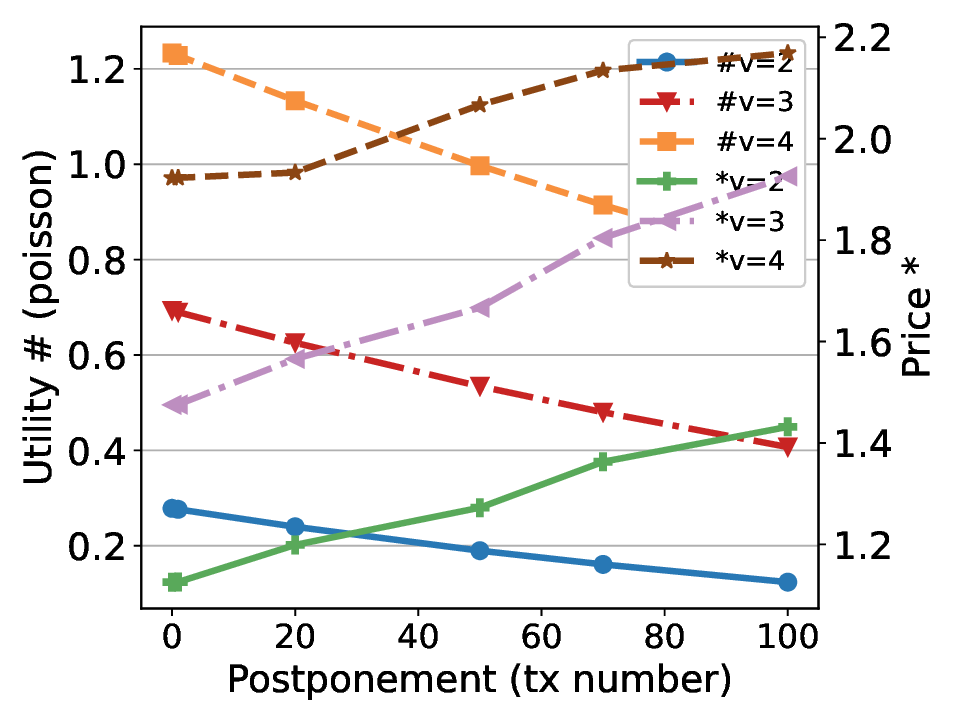}\\
			\caption{Utility and transaction fee of FBR in different postponements with Poisson arrival.}
			\label{fig:FBRpoisson}
		\end{minipage}
  \vspace{-0.5cm}
	\end{figure}

The probability of being incorporated in the next valid block is also an important metric for the SP user. As illustrated in Fig. \ref{fig:successlinear} and Fig. \ref{fig:successpoisson}, there is a positive correlation between the transaction value and its success probability. In the Bitcoin-like PoW system, the success probability diminishes as the transaction arrival time gets later. Whereas in the Ethereum-like PoS system, the success probability tends to increase.

\begin{observation}
  \emph{
  Postponing broadcasting transactions when the block generating interval is exponentially distributed (resp. fixed) causes expected utilities to decrease (resp. increases) for the SP user with \textbf{FBR} strategy. 
  }
\end{observation}

We selected a set of mempool states, consisting of 80 transactions within the present mempool, to evaluate the impact of the SP user postponing the broadcasting of transactions in the Bitcoin-like PoW system. Fig. \ref{fig:FBRlinear} and Fig. \ref{fig:FBRpoisson} depict the results under linear transaction arrival and Poisson transaction arrival. 
The horizontal axis in the graph denotes the count of transactions generated between the generation and broadcasting time of the SP transaction, which is denoted as the postponement. It also represents the number of transactions that the SP user can observe beyond the SP transaction generation time. The left vertical axis corresponds to the utility acquired by the SP user through the FBR strategy under different broadcasting time, while the right vertical axis signifies the optimal transaction fee established by the SP user.
In the Bitcoin-like PoW system, a noticeable reduction in the utility of the SP user can be observed as the postponement increases. This decrease is accompanied by a significant increase in the optimal transaction 
 fee that the SP user needs to set. 

In the Ethereum-like PoS system, when the SP user's values are 2, 3, and 4, and the other transactions arrive at a linear rate, the utilities that the SP user can obtain by choosing to broadcast transaction at time $\bar{T}$ are $0.215, 1.214, 2.212$ respectively. They are all significantly higher than the utility of broadcasting the transaction immediately. 
Similarly, when the transaction arrival process is a Poisson process, the utilities of the SP user are 0.214, 1.212 and 2.211 respectively, which are still higher than the utility of broadcasting at time $t_S$ when $t_S$ is less than $\bar{T}$.

\subsection{Experimental results of semi-strategic seniors}

\begin{observation}
    \emph{The utility of the SP user increases as the frequency $\gamma_S$ increases when other users are semi-strategic and the block generating interval is exponentially distributed.}
\end{observation}

\begin{figure}[htb]
 \vspace{-0.3cm}
		\begin{minipage}{0.49\linewidth}
			\setlength\abovecaptionskip{0.5pt}
			\setlength\belowcaptionskip{-1pt}
			\centering
			\includegraphics[width=0.9\textwidth]{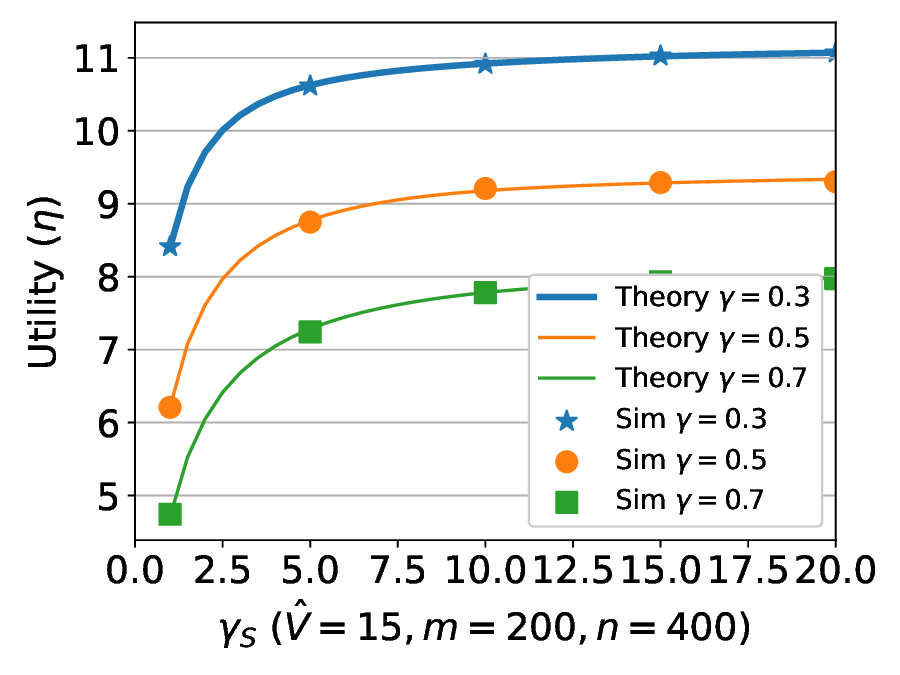}\\
			\caption{Utility under different $\gamma$.}
			\label{fig:differentgamma}
		\end{minipage}
		\begin{minipage}{0.49\linewidth}
			\setlength\abovecaptionskip{0.5pt}
			\setlength\belowcaptionskip{-1pt}
			\centering
			\includegraphics[width=0.9\textwidth]{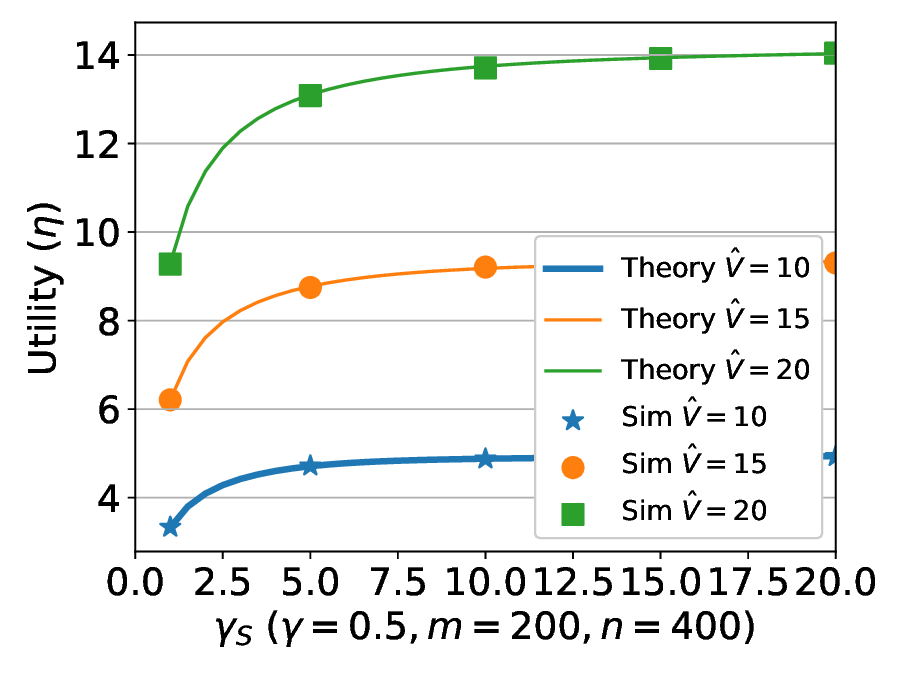}\\
			\caption{Utility under different $\hat{V}$.}
			\label{fig:differentvalue}
		\end{minipage}
  		\begin{minipage}{0.49\linewidth}
			\setlength\abovecaptionskip{0.5pt}
			\setlength\belowcaptionskip{-1pt}
			\centering
			\includegraphics[width=0.9\textwidth]{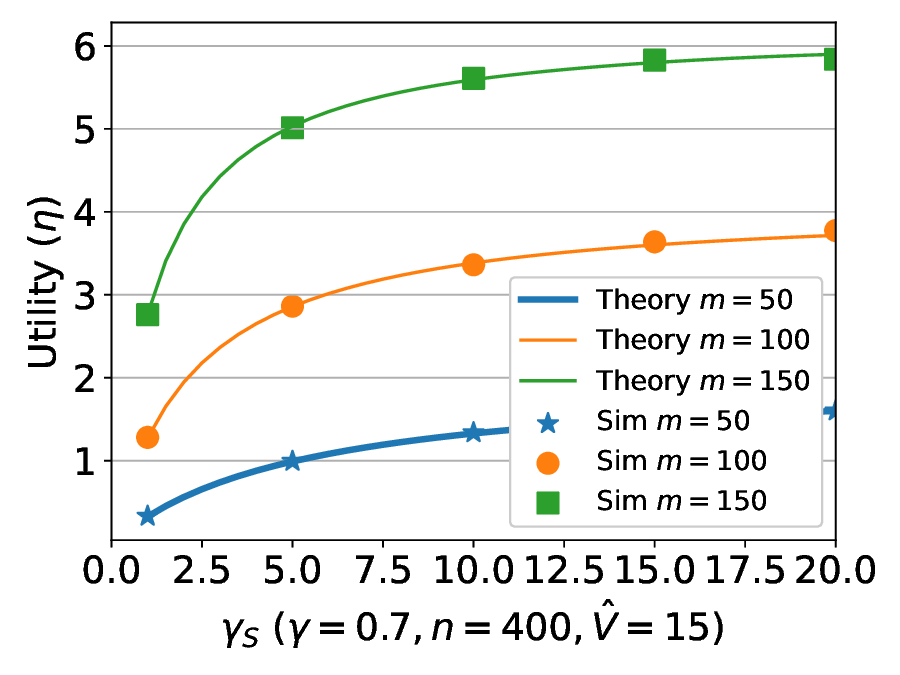}\\
			\caption{Utility under different $m$.}
			\label{fig:differentm}
		\end{minipage}
		\begin{minipage}{0.49\linewidth}
			\setlength\abovecaptionskip{0.5pt}
			\setlength\belowcaptionskip{-1pt}
			\centering
			\includegraphics[width=0.9\textwidth]{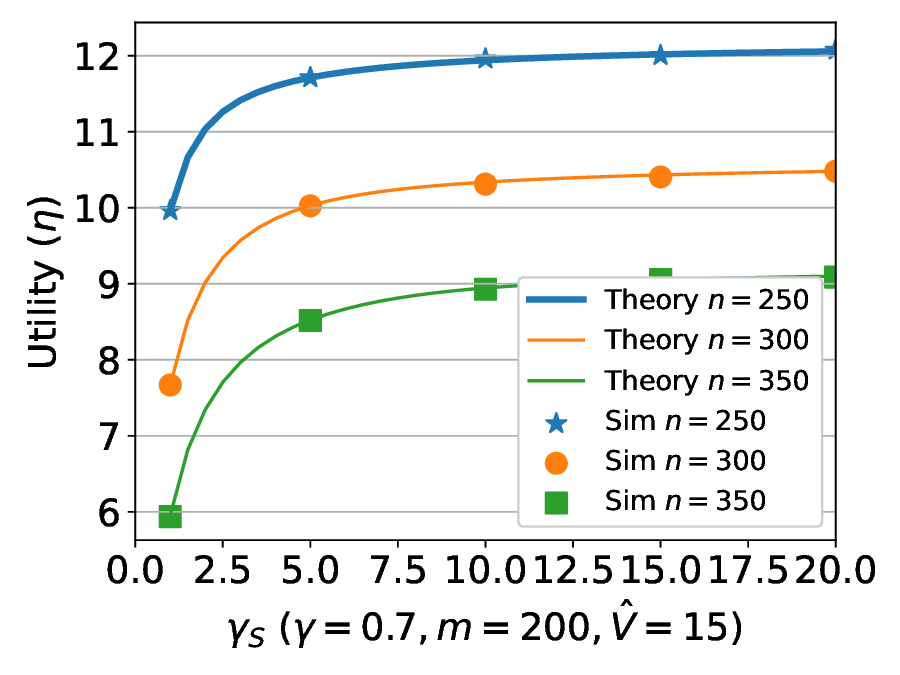}\\
			\caption{Utility under different $n$.}
			\label{fig:differentn}
		\end{minipage}
  \vspace{-0.3cm}
	\end{figure}

We explore the utility of the SP user per transaction at different frequencies of fee bumping when other users are semi-strategic.
Fig. \ref{fig:differentgamma} $\sim$ Fig. \ref{fig:differentn} show the utility of each transaction for the SP user under different parameters. The simulation results are consistent with the theoretical results in all cases, which verifies the correctness of our model. It can be observed that the utility of the SP user always increases with the increase of $\gamma_S$.
The aforementioned observation confirms that in the Bitcoin-like PoW system, the optimal strategy for the SP user is to increase the transaction fee immediately when his transaction fee is less than the fee threshold.
At the same time, the utility of the SP user increases with the increase of $V$ and $m$, and decreases with the increase of $\gamma$ and $n$. 

\begin{figure}
\vspace{-0.3cm}
  		\begin{minipage}[htb]{0.49\linewidth}
			\setlength\abovecaptionskip{0.5pt}
			\setlength\belowcaptionskip{-1pt}
			\centering
			\includegraphics[width=0.9\textwidth]{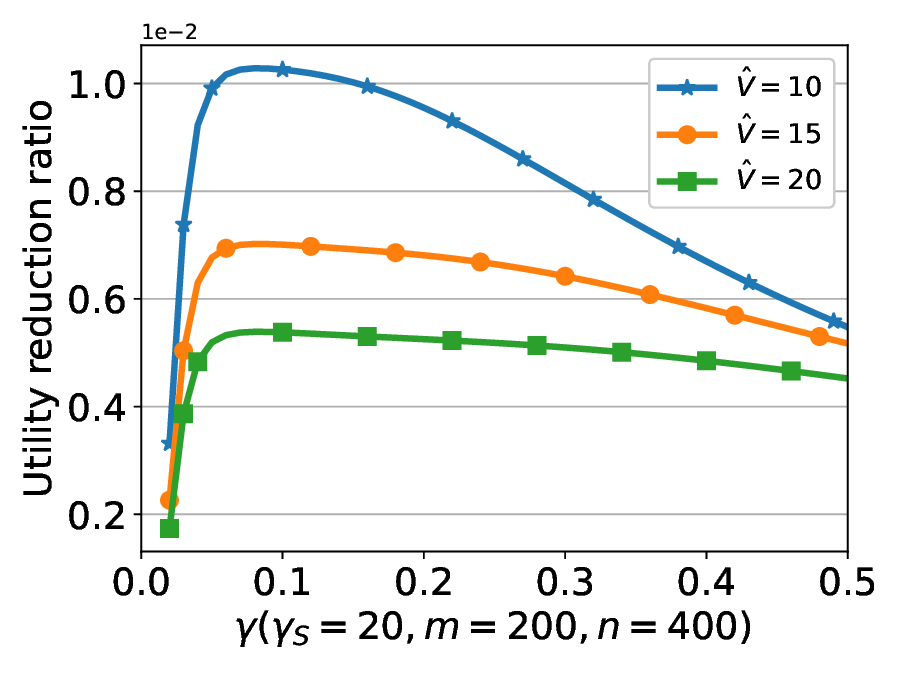}\\
			\caption{Utility reduction ratio under different $\gamma$ when $\gamma_S=20$.}
			\label{fig:gammas20}
		\end{minipage}
  		\begin{minipage}[htb]{0.49\linewidth}
			\setlength\abovecaptionskip{0.5pt}
			\setlength\belowcaptionskip{-1pt}
			\centering
			\includegraphics[width=0.9\textwidth]{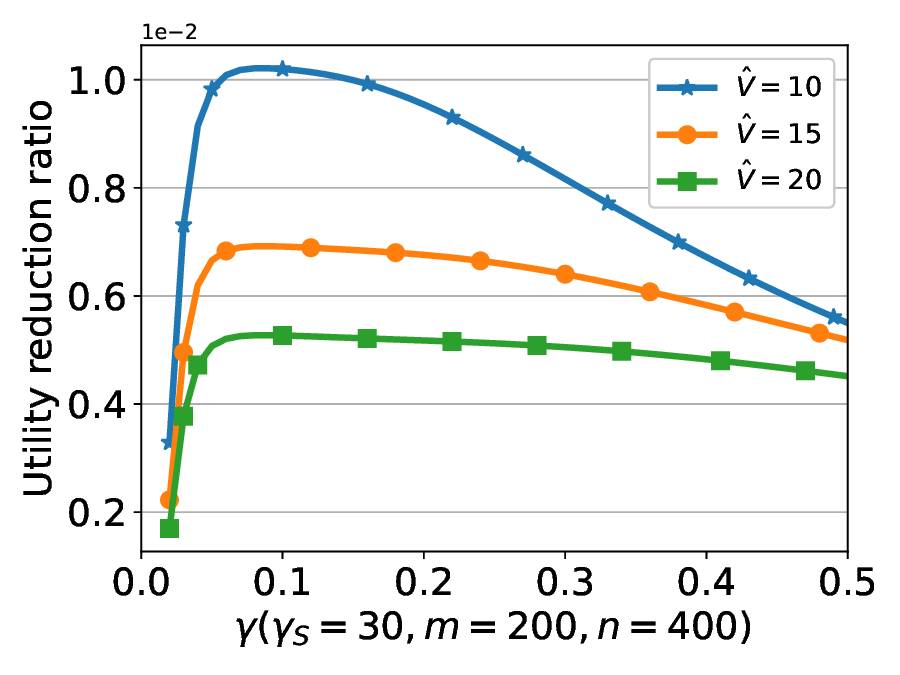}\\
			\caption{Utility reduction ratio under different $\gamma$ when $\gamma_S=30$.}
			\label{fig:gammas30}
		\end{minipage}
  \vspace{-0.7cm}
	\end{figure}

We further investigate the influence of other users' strategies on the utility of the SP user and present the findings in Fig. \ref{fig:gammas20} and Fig. \ref{fig:gammas30}. The utility reduction ratio refers to the proportion of utility reduction among the SP user in relation to $V$. 
As $\gamma$ increases, the utility reduction ratio exhibits an initial increase followed by a subsequent decrease. Consequently, we can claim that the SP user exhibits higher sensitivity to other users' strategy when $\gamma$ resides within an appropriate range. 

%% file: section/related_work.tex
\section{related work}
\label{sec:related work}

In what follows, we describe the recent progress on the waiting time analysis and transaction price prediction.

\emph{Waiting time analysis.} 
Geissler et al. \cite{ref:discrete time analysis} developed a discrete-time queuing model to analyze the distribution of transaction waiting time without priorities. Considering the priority of the transaction based on transaction fee or other parameters, Kasahara \cite{ref:bernoulli exponential2} and Kawase et al. \cite{ref:priority queueing analysis} modeled the transaction confirmation process of Bitcoin as a priority queueing system with batch services and derived the average transaction confirmation time.
  Fiz et al. \cite{ref:confirmation delay prediction} developed a model which adopts a supervised classification approach to predict the likelihood of an unconfirmed transaction being processed in the subsequent block. 
Zhang et al.\cite{ref:a_classification_view} considered the prediction of transaction confirmation as a classification problem, which predicts the confirmation time as predicting a specific time interval rather than a timestamp.

\emph{Transaction fee prediction.} 
Al-Shehabi et al. \cite{ref:al-shehabi} proposed a model to predict the minimum transaction fee required for users to process transactions within the target block in the Bitcoin system using the perceptron machine learning classification algorithm. Mars et al. \cite{ref:mars ethereum} presented a method utilizing deep learning models (LSTM and GRU) and the Prophet model to predict gas prices in the next Ethereum block, which achieved a significant decrease in mean squared error (MSE).
Liu et al. \cite{ref:liu2020effective} proposed an algorithm based on machine learning regression (MLR) considering block gas limits to predict the minimum gas price in the next block, which achieved an accuracy of 74.9\%.

\section{Conclusion}
\label{section:conclusion}
This paper primarily investigates the transaction fee strategy of the strategic user, focusing on transaction broadcasting time and transaction fee, while considering the usage of fee bumping by other users in various blockchain systems. We find that in fee bumping absence, the strategic user in the PoW (Bitcoin-like) system should immediately broadcast transactions, while in the PoS (Ethereum-like) system should broadcast just before the block generation. When fee bumping is used by other users and block generating interval is exponentially distributed, the strategic user should promptly increase fees when below the mempool threshold. This conclusion is supported by our continuous-time Markov chain model. 